\newcommand{\sartre}{Sar\emph{t}re}
\documentclass[Physsubmission, Phys]{SciPost}

\hypersetup{
    colorlinks,
    linkcolor={red!50!black},
    citecolor={blue!50!black},
    urlcolor={blue!80!black}
}

\usepackage[bitstream-charter]{mathdesign}
\DeclareSymbolFont{usualmathcal}{OMS}{cmsy}{m}{n}
\DeclareSymbolFontAlphabet{\mathcal}{usualmathcal}

\begin{document}

\begin{center}{\Large \textbf{
Subnucleon fluctuations in coherent and incoherent ultra-peripheral AA collisions at LHC and RHIC with the Sar\emph{t}re
event generator.
\\
}}\end{center}

\begin{center}
Tobias Toll
\end{center}

\begin{center}
Indian Institute of Technology Delhi\\
tobiastoll@iitd.ac.in
\end{center}

\begin{center}
\today
\end{center}


\definecolor{palegray}{gray}{0.95}
\begin{center}
\colorbox{palegray}{
  \begin{tabular}{rr}
  \begin{minipage}{0.1\textwidth}
    \includegraphics[width=22mm]{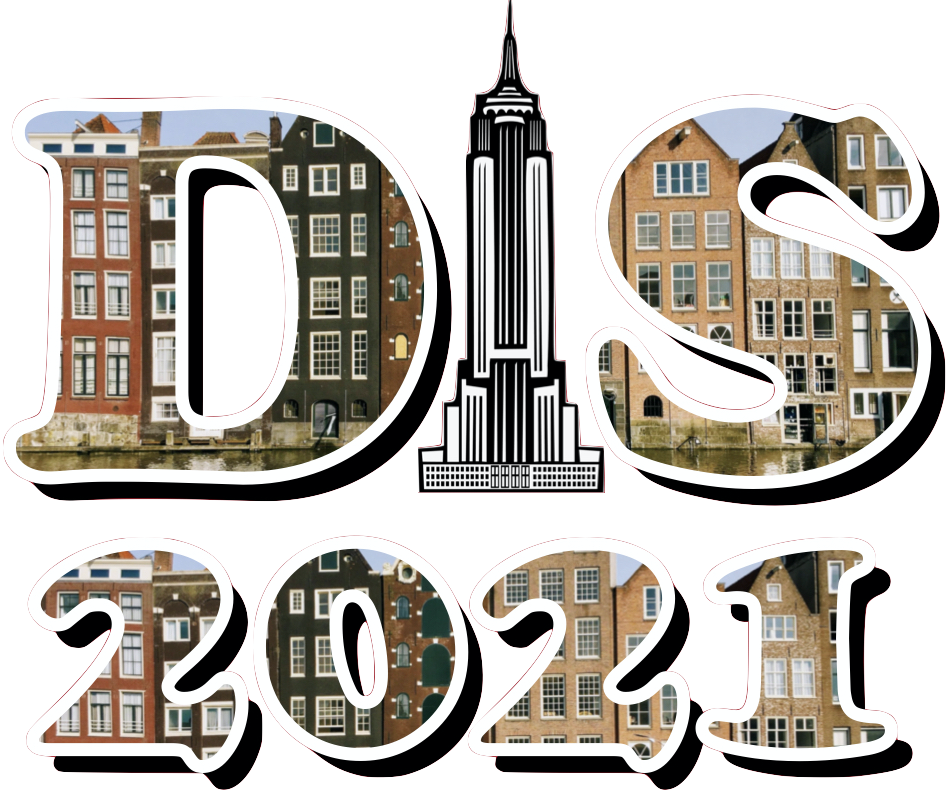}
  \end{minipage}
  &
  \begin{minipage}{0.75\textwidth}
    \begin{center}
    {\it Proceedings for the XXVIII International Workshop\\ on Deep-Inelastic Scattering and
Related Subjects,}\\
    {\it Stony Brook University, New York, USA, 12-16 April 2021} \\
    \doi{10.21468/SciPostPhysProc.?}\\
    \end{center}
  \end{minipage}
\end{tabular}
}
\end{center}

\section*{Abstract}
{\bf
\sartre~has been extensively used for describing photon-nuclei processes at the electron-ion collider (EIC) as well as ultra-peripheral collisions (UPC) at LHC and RHIC. \sartre~is an event generator which implements the dipole model for DIS, and models the transverse geometry of the target nucleus or proton in coordinate space. It uses the Good-Walker mechanism for simulating fluctuations which contribute to the incoherent cross section for which the target breaks up after the interaction. With improved precision of UPC measurements in the last years, a detailed test of the dipole model has become possible, and \sartre's model was found lacking. In these proceedings we add subnucleon fluctuations to the nucleus and show that this is sufficient for describing the vast majority of the present measurements. We also find that for larger momentum transfers in the nucleus, which probes gluon fluctuations at higher resolution, the current complexity of the model may not suffice. Future measurements at the LHC, RHIC and especially the EIC has the potential to reveal these gluon vacuum fluctuations and glean novel insights into the self-interacting quantum field of QCD. 
}



\section{Introduction}
\label{sec:intro}
\begin{figure}[h]
\centering
\includegraphics[width=0.3\textwidth]{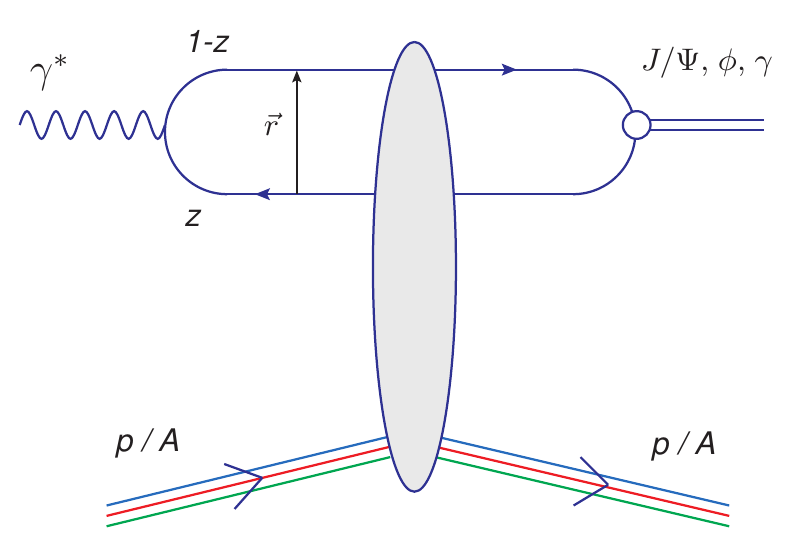}
\caption{In the dipole model  a virtual photon fluctuates into a $q\bar q$ state which interacts strongly with the proton. In exclusive diffraction the dipole subsequently forms a vector meson.}
\label{fig:dipole}
\end{figure}

The Monte Carlo event generator \sartre~ \cite{Toll:2012mb, Toll:2013gda, Sambasivam:2019gdd} was the first model based on event-by-event fluctuations to implement exclusive incoherent diffraction for heavy nucleus targets in deeply inelastic scattering (DIS) events. \sartre~ uses the bSat dipole model \cite{GolecBiernat:1998js, GolecBiernat:1999qd, Kowalski:2003hm, Kowalski:2006hc, Rezaeian:2012ji, Mantysaari:2018nng} for which the amplitude is given by \cite{Kowalski:2003hm}:
 \begin{eqnarray}
    	\mathcal{A}_{T,L}^{\gamma^* p \rightarrow J/\Psi p} (x_{\mathbb{P}},Q^2,\textbf{$\Delta$})=
	i\int {\rm d}^2 \textbf{r}\int {\rm d}^2\textbf{b}\int \frac{{\rm d}z}{4 \pi}\times (\Psi^*\Psi_V)_{T,L}(Q^2,\textbf{r},z)e^{-i[\textbf{b}-(1-z)\textbf{r}].\textbf{$\Delta$}} \frac{d\sigma _{q\bar{q}}}{d^2\textbf{b}}(\textbf{b},\textbf{r},x_\mathbb{P})
	\label{eq:amp}
 \end{eqnarray}
This amplitude describes the interaction of a virtual photon, which fluctuates into a quark-anti quark dipole which interacts via one or many colourless two-gluon exchanges with the proton, and then recombines into a vector meson as depicted in Fig. \ref{fig:dipole}. Here, $T$ and $L$ represent the transverse and longitudinal polarisation of the virtual photon $\gamma^*$, \textbf{r} is the transverse size of the dipole, \textbf{b} is the impact parameter of the dipole relative to the proton, $z$ is the energy fraction of the photon taken by the quark and $\bf{\Delta} \equiv \sqrt{-t} $ is the transverse part of the four-momentum difference of the outgoing and the incoming proton. $(\Psi^* \Psi_V)$ denotes the wave-function overlap between the virtual photon and the produced vector meson, which is taken to be a boosted Gaussian \cite{Mantysaari:2018nng}. 
The dipole cross section $\frac{d\sigma _{q\bar{q}}}{d^2\textbf{b}}(\textbf{b},\textbf{r},x_\mathbb{P})$ describe the strong interaction. 

The bSat dipole cross-section is given by \cite{Bartels:2002cj}:
 \begin{eqnarray}
    	\frac{d\sigma _{q\bar{q}}}{d^2\textbf{b}}(\textbf{b},\textbf{r},x_\mathbb{P})=
	2\bigg[1-\text{exp}\bigg(-
\frac{\pi^2}{2N_C} \textbf{r}^2 \alpha_s(\mu^2) x_\mathbb{P} g(x_\mathbb{P},\mu^2)	T_p(\textbf{b})\bigg)\bigg].
\end{eqnarray}
A few things are worthy of note in this expression. The gluon density is described in momentum space in the longitudinal direction and coordinate space in the transverse directions, as the Bjorken $x$ variable is the longitudinal momentum fraction of the gluons that enter in the gluon density $xg(x)$. The transverse gluon density profile is given by the Gaussian $T_p(\textbf{b}) = \frac{1}{2 \pi B_G}\exp\bigg(-\frac{\textbf{b}^2}{2B_G}\bigg)$. In this way the dipole model becomes an excellent tool for investigating spatial gluon distributions and fluctuations in coordinate space. Furthermore, the dipole cross-section saturates for large dipole sizes and large gluon densities. This is due to non-linear effects emmenating from multiple two-gluon exchanges. 

\subsection{Fluctuations}
In diffraction there are two classes of events. Firstly, \emph{coherent} scattering in which the nucleic target remains intact in the interaction. This is possible as the two-gluon exchanges do not carry any quantum numbers. Secondly, \emph{incoherent} scattering in which the nucleus becomes excited in the interaction and subsequently de-excites through the emission of a photon, one or many protons or neutrons or, if the excitation energy is large enough, the nucleus may shatter into fragments. This classification is described in the Good-Walker picture \cite{Good:1960ba} in terms of fluctuations. Here, the coherent cross section is described by the event-by-event average of the amplitude due to some variation $\Omega$, and the incoherent cross section by its variation:
    \begin{eqnarray}
    	\frac{\rm{d} \sigma_{\rm coherent}}{{\rm d}t} &=& \frac{1}{16 \pi} \big| \left<\mathcal{A}(x_{\mathbb{P}},Q^2,\textbf{$\Delta$})\right>_\Omega\big|^2 \nonumber \\
    	\frac{{\rm d} \sigma_{\rm incoherent}}{{\rm d}t} &=& \frac{1}{16 \pi}\bigg(\big< \big| \mathcal{A}(x_{\mathbb{P}},Q^2,\textbf{$\Delta$})\big|^2\big>_\Omega \big| -\big<\mathcal{A}(x_{\mathbb{P}},Q^2,\textbf{$\Delta$})\big>_\Omega\big|^2\bigg)
    	\end{eqnarray}

For nuclei, this formalism was introduced in \sartre~ through variations in the configurations of nucleons in a nucleus following a Woods-Saxon distribution \cite{Toll:2012mb}. This formalism has been used as a bench mark for designing the interaction region of the electron-ion collider (EIC) with the goal to measure both the coherent and incoherent cross-sections in $e$A collisions. Only lately has the model been confronted with measurements from the LHC and RHIC, which measure the incoherent and coherent cross-sections in ultra-peripheral collisions (UPC) in which the colliding nuclei interact electromagnetically via exchange of a pseudo-real photon. This study was presented in \cite{Sambasivam:2019gdd}. There we found that \sartre~describes the coherent cross section within experimental errors but underpredicts the incoherent cross sections. This is an indication that the model of the incoherent cross section did not contain enough fluctuations. 

\begin{figure}[h]
\centering
\includegraphics[width=0.99\textwidth]{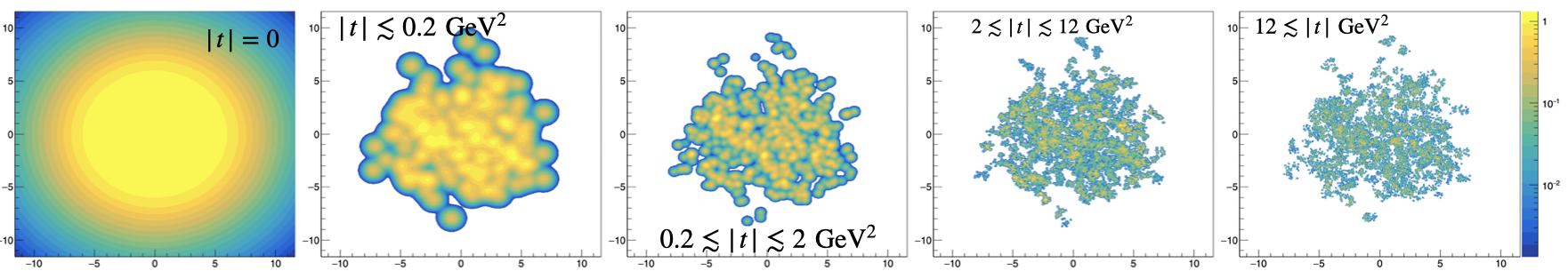}
\caption{The gluonic structure of a lead nucleus at increasing values of $|t|$ with parameters from \cite{Kumar:2021zbn}. Units are in fm.}
\label{fig:nucleus}
\end{figure}

M\"antysaari and Schenke have shown that in incoherent $ep$ scattering at HERA, one can describe the cross section with the dipole model if, inside the proton, there are three Gaussian hotspots whose position vary event-by-event \cite{Mantysaari:2017dwh, Mantysaari:2016ykx,Mantysaari:2016jaz}. The transverse gluon thickness of the proton then becomes:
\begin{eqnarray}
    	T_p(\textbf{b}) = \frac{1}{N_q}\sum_{i=1}^{N_q}T_q(\textbf{b-b$_i$}),~~~~~
	T_{q}(\textbf{b}) = \frac{1}{2 \pi B_{q}}\exp\big[-\frac{\textbf{b}^2}{2B_{q}}\big]
\end{eqnarray}
Here $N_q$ = 3 and $\textbf{b}_i$ are the locations of hotspots sampled from a Gaussian width $B_{qc}$ and $B_q$ is the width of the hotspots.  $B_{qc}$ and $B_q$ control the amount of fluctuations in the proton geometry at low momentum transfer and are constrained by the coherent and incoherent data. As can be seen in eq.~\eqref{eq:amp}, the amplitude is a Fourier transform between momentum transfer $\Delta=\sqrt{-t}$ and impact parameter $b$. This means that large values of $|t|$ correspond to higher resolution in $b$. Therefore, we expect the small-size effects of intra-nuclear fluctuations to become important at larger values of $|t|$. In a recent paper \cite{Kumar:2021zbn} we show that the entire measured incoherent $|t|$ spectrum at HERA can be described with a model of hotspots within hotspots within hotspots. We further discovered that the sizes and number of hotspots at each level of substructure are highly correlated hinting that there is a spatial scaling in the gluonic fluctuation substructure of the proton, with the caveat that our model does not contain gluon-gluon correlations. An illustration of this for a lead nucleus is depicted in Fig.~\ref{fig:nucleus}.

\section{Results}
In these proceedings we confront the hotspot model, implemented in \sartre, with recent UPC measurements with heavy nuclei at LHC and RHIC using two levels of substructure in the nucleus: nucleons and one level of hotspots within them. For this purpose we set the center-of-mass of the hotspots at the center of the nucleon, and center-of-mass of the nucleon configuration at the center of the nucleus. All the dipole parameters are described in \cite{Sambasivam:2019gdd}. For the hotspots, we use $B_{qc}=4.2$~GeV$^{-2}$, $B_q=0.8$~GeV$^{-2}$. Following \cite{Mantysaari:2017dwh} we also implement saturation scale fluctuations with $\sigma=0.75$.

\begin{figure}[h]
\centering
\includegraphics[width=0.35\textwidth]{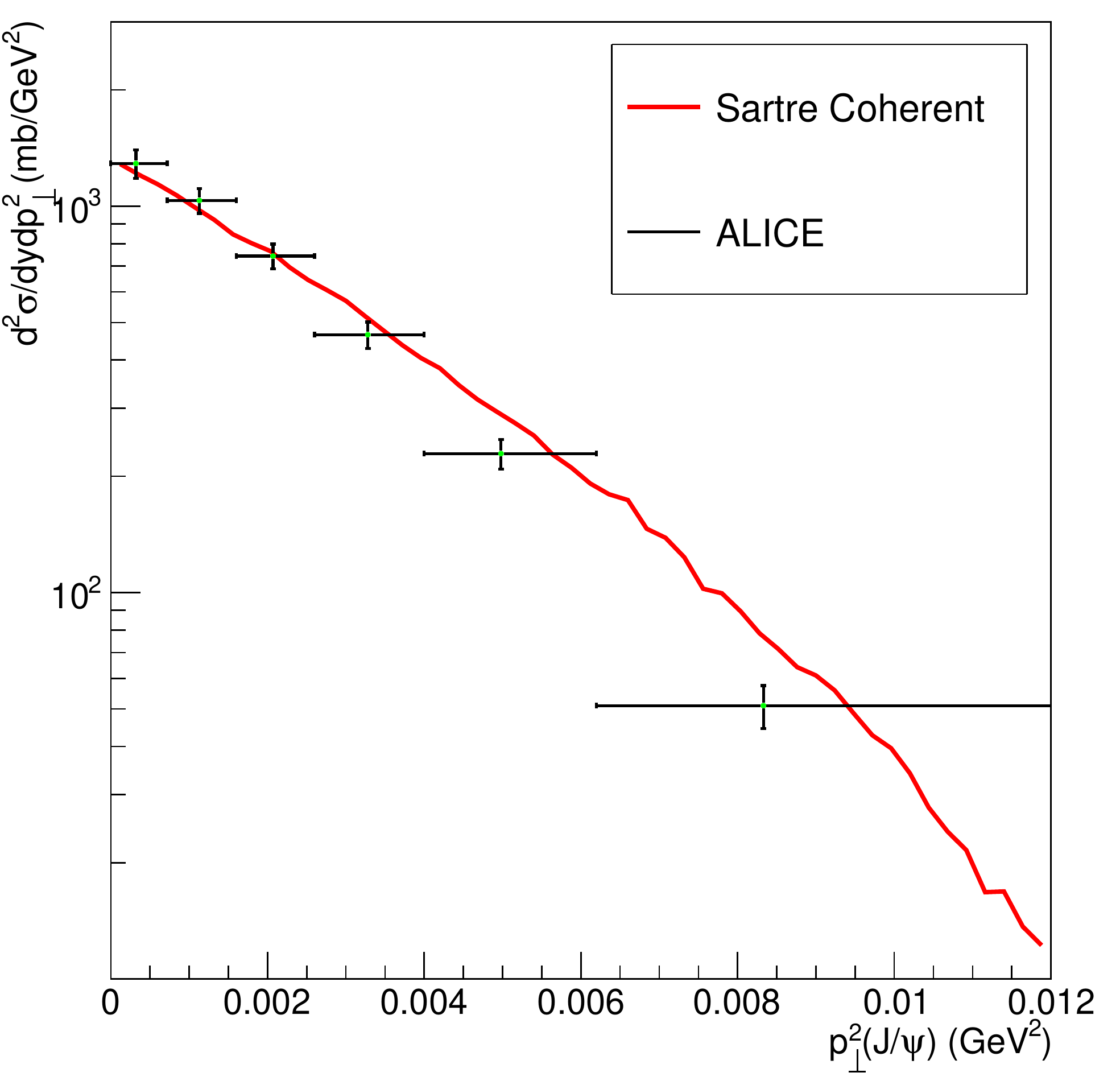}
\includegraphics[width=0.35\textwidth]{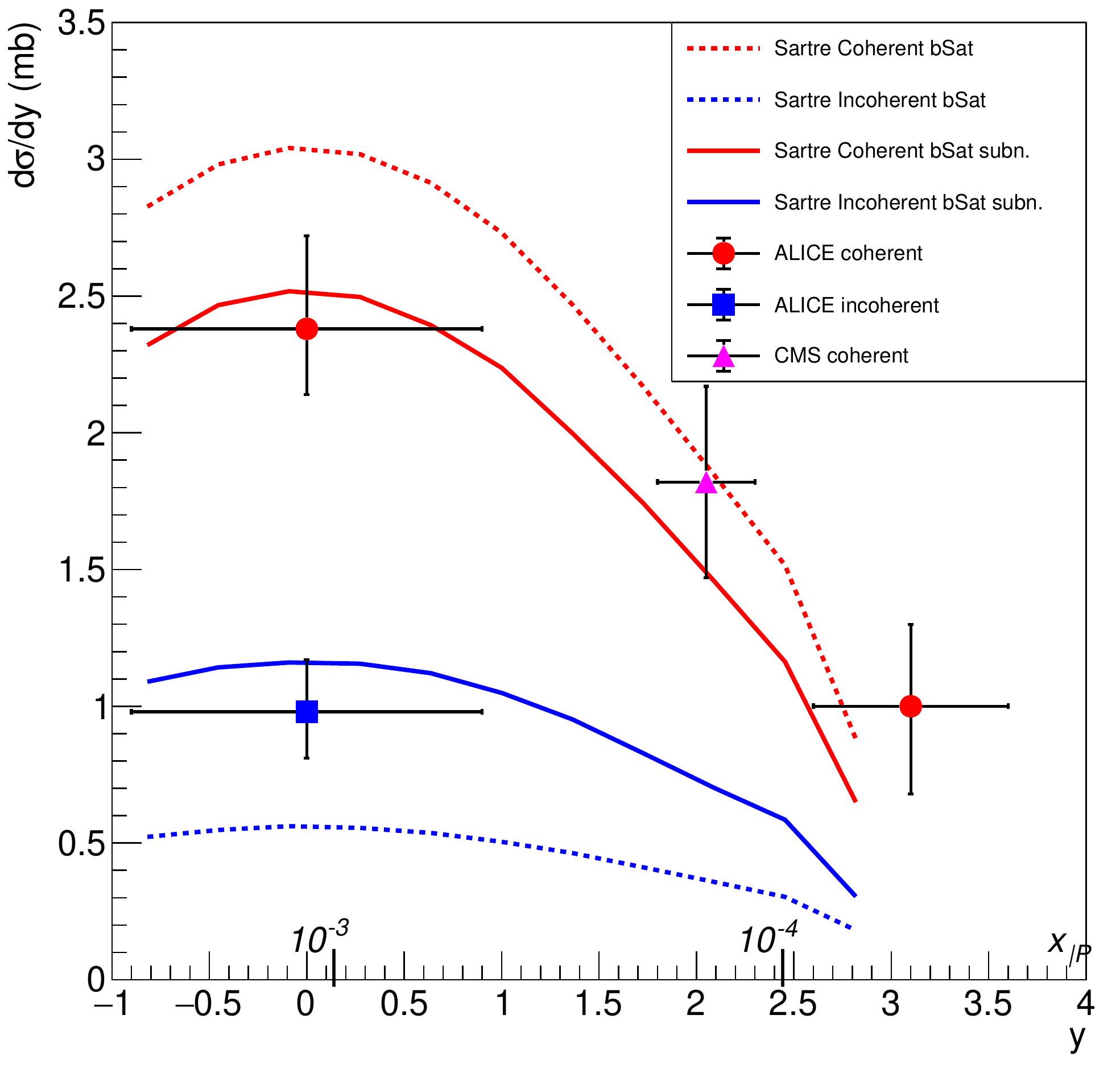}
\caption{Left: A comparison with coherent Pb-Pb UPC from ALICE collaboration at $\sqrt{s}=5.02~$TeV \cite{ALICE:2021tyx} with \sartre. Right: A comparison with coherent and incoherent data in Pb-Pb collisions from the ALICE \cite{SCAPPARONE:2013isa} and CMS \cite{Khachatryan:2016qhq} collaborations compared to \sartre~ with and without subnuclear fluctuations.}
\label{fig:ALICE}
\end{figure}

In Fig.~\ref{fig:ALICE}, left panel, we compare \sartre~with the coherent $t$-spectrum in Pb+Pb$\rightarrow$ Pb+$J/\Psi$+Pb events at $\sqrt{s}=5.02~$TeV measured by the ALICE collaboration \cite{ALICE:2021tyx}. We see that \sartre~describes the data well within uncertainties. In the right panel we compare the rapidity spectrum for the same process at $\sqrt{s}=2.76$~TeV of both incoherent and coherent cross sections. Here we compare the measurements with two versions of \sartre: with and without subnuclear fluctuations. One should note that this data has been integrated over all $t$ and the result will be somewhat sensitive to large $t$ effects in the fluctuation spectrum. We see that a good description is found when subnucleon fluctuations are included in the model. 

\begin{figure}[h]
\centering
\includegraphics[width=0.45\textwidth]{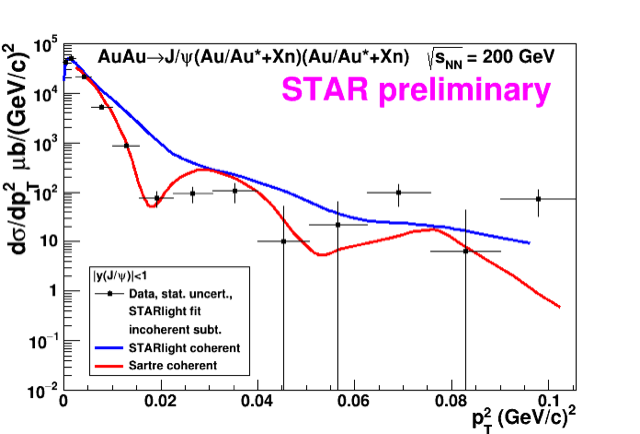}
\includegraphics[width=0.45\textwidth]{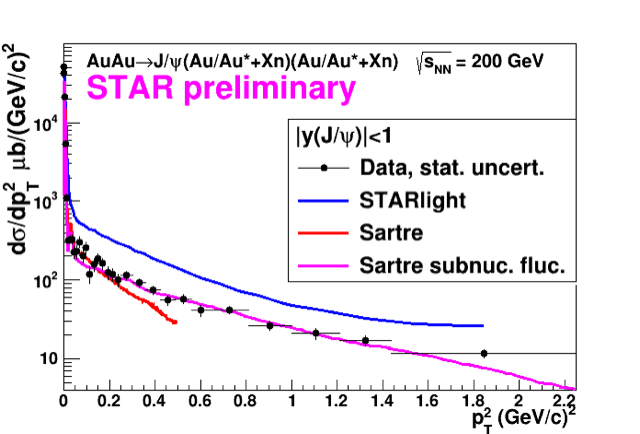}
\caption{Left: A comparison with coherent Au-Au UPC from the STAR collaboration at $\sqrt{s}=200~$GeV with \sartre. Right: A comparison with coherent and incoherent preliminary data in Au-Au collisions from the STAR collaboration compared to \sartre~ with and without subnuclear fluctuations.}
\label{fig:STAR}
\end{figure}

In Fig.~\ref{fig:STAR} we confront \sartre~with exclusive $J/\Psi$ measured in Au-Au collision at STAR as a function of $p_t^2\approx |t|$, where $p_t$ is the transverse momentum of the vector meson. In the left panel \sartre~describes the coherent measurement well within uncertainties. In the right panel, we see that \sartre~without subnucleon fluctuations is unable to describe the data for $p_t^2>0.25$~GeV$^2$, but with one level of subnucleon fluctuations we are able to describe the data well up to $p_T^2\sim 2.2$~GeV$^2$. This is consistent with our findings from $ep$ described in \cite{Kumar:2021zbn} where a further substructure needed to be introduced to the nucleon in order to describe the data for $|t|>2.5$~GeV$^2$. 

\section{Conclusion}
We have shown that subnucleon fluctuations are needed to describe the recent AA UPC measurements from ALICE, CMS, and STAR. For observables which integrate over $|t|$ the large $|t|$ tails created by subnucleon fluctuations give a significant contribution as can be seen in the rapidity spectrum for both coherent and incoherent scattering. The good description of the coherent $|t|$ spectrum at small $|t|$ gives further confidence in this conclusion. The $p_t^2$ spectrum measured by STAR shows that the contribution from subnucleon fluctuations becomes significant for $p_t^2>0.25$~GeV$^2$. The studies of incoherent scattering in $ep$ from HERA suggest that for $p_T^2\gtrsim 2$~GeV$^2$ the hotspot model as implemented will no longer describe the data well and additional substructure then becomes necessary. These measurements on UPC should become available in the very near future. It appears that the cross section at this point is too small to give a significant contributions to other observables within uncertainties. The subnuclear fluctuations are implemented in \sartre~for UPC, and they will be implemented for $e$A collisions as well in the near future.

The EIC will be able to measure the incoherent $|t|$ spectrum with unprecedented reach and precision for both $ep$ and $e$A collisions, and will thus be able to measure the structure of the gluon fluctuations in nuclear matter in a way which will give new insights into self-interacting quantum field theories otherwise not available to us.

\section*{Acknowledgements}
This work was done in collaboration with Arjun Kumar. We thank the physics department of IIT Delhi for support.



\bibliography{bibliography.bib}

\nolinenumbers

\end{document}